\documentclass[10pt]{revtex4}
\usepackage{hyperref}
\begin{document}

\title{ A superspace description of Chern-Simons theory  in Batalin-Vilkovisky formulation}
 \author{ Sudhaker Upadhyay} \email {  sudhakerupadhyay@gmail.com} 
 \address{Departamento de F\'{\i }sica Te\'{o}rica, Instituto de F\'{\i }sica, UERJ - 
 Universidade do Estado do Rio de Janeiro,
 \\ \small \textnormal{ \it Rua S\~{a}o Francisco Xavier 524, 20550-013 Maracan\~{a}, Rio de 
 Janeiro, Brasil}\normalsize.}
 \author{ Manoj Kumar Dwivedi}\email{manojdwivedi84@gmail.com}
\affiliation { Department of Physics, 
Banaras Hindu University, 
Varanasi-221005, INDIA.  }

\author{ Bhabani Prasad Mandal}\email{ bhabani.mandal@gmail.com}
\affiliation { Department of Physics, 
Banaras Hindu University, 
Varanasi-221005, INDIA.  }
\begin{abstract}
We discuss the extended BRST and anti-BRST symmetry (including shift symmetry) in the  
Batalin-Vilkovisky 
(BV) formulation for the Chern-Simons (CS) theories in $(2+1)$ spacetime dimensions. Further 
we develop the superspace 
description of BV formulation for such theories. Interestingly,  the extended BRST invariant   
CS theories can be described in superspace in  covariant manner  
with the help of one more (Grassmann) coordinate. 
However, a superspace with two Grassmann coordinates are required for a manifestly covariant 
formulation 
of the extended BRST and extended anti-BRST invariant   actions for these theories.
\end{abstract}
\maketitle
\section{Introduction}
In recent past years, the CS gauge theories have been studied considerably with great 
interests \cite{as,ew,ew1,town,jf,db,fd}. 
 The $(2+1)$ dimensional  CS theories with compact gauge group give natural explanations
 \cite{ew} for many constructions in conformal field theory and integrable
lattice models that have been intensively studied \cite{ew1}.
The CS theories also get relevance  in anti-de-Sitter (adS) supergravity theories \cite{town}.
The CS theories, quantized  particularly in 
axial gauge,  obey  the topological supersymmetry which was known to hold in the Landau
(covariant) gauge \cite{db, fd}. Topological field theories are a class of gauge models with 
the peculiarity that their
observables are of topological nature, as for instance the knot and link invariants in the
case of $(2+1)$-dimensional CS theories \cite{ew}. The axial gauge has peculiar 
interest for the CS theories and other topological field models \cite{ew,fd1, ab} since in 
this gauge the finiteness problem is obvious due to the complete absence of radiative 
corrections.  The Green functions of the CS theories quantized in axial gauge are shown to be 
calculable as the unique exact solutions of the Ward identities which express 
the invariance of the theory under topological supersymmetry \cite{fd,kla}. Another important 
feature 
of topological supersymmetry, which makes it physically relevant, is its role in the 
construction
of observables \cite{eg}. The intriguing point of the CS model   is 
the existence
of a very large algebra of symmetries.

The  BV  approach \cite{bv,bv1,wei} is the most  powerful
quantization algorithm presently available which allows us to deal with very general gauge 
theories, including those with open or reducible gauge symmetry algebras.
The BV method also provides a convenient way of analysing the possible violations of 
symmetries
of the action by quantum effects \cite{wei}.
The  BV formalism,also known as field antifield formulation(previously and independently 
introduced by
Zinn-Justin \cite{zinn}) generalizes the BRST approach \cite{brs,brs1}. It is usually used as a 
covariant method to perform the gauge-fixing
in quantum field theory, but was also applies to other problems like analysing
possible deformations of the action and anomalies.

 A superspace description \cite{jm, jm1, jm2} for the
non-anomalous  gauge theories in BV formulation has been studied extensively \cite{ad,ad1,ad2, 
ba}.
It has been shown  that the extended BRST and extended
anti-BRST invariant actions of these theories
(including some shift symmetry) in   BV  formulation  \cite{ad,ba,fk,subm,sudh}, yield 
naturally  the proper identification of the antifields through equations of motion.
The shift symmetry plays an important role and gets relevance, for instance,  in  inflation 
particularly in supergravity \cite{brax} as well as in Standard Model \cite{heb}.  However, in 
usual BV formulation the  antifields  are calculated from the expression of gauge-fixing 
fermion. Recently, this formulation 
has been extended for the theory of perturbative gravity \cite{sud}. We  extend such 
formulation 
to the topological gauge theory in $(2+1)$ dimensions. 

In the present work we attempt to provide a superspace version of CS theory
in BV formulation. For this purpose, we first consider BRST invariant 
CS theory  in axial and Landau gauges. Then we extend the BRST symmetry of the theory by 
including 
shift symmetry. The advantage of making such analysis is that the antifields get
their own identifications naturally. Further, we describe the extended BRST invariant CS 
theory
in superspace using only one Grassmann coordinate together with
$(2+1)$ spacetime dimensions. However, for both extended BRST  and extended 
anti-BRST invariant CS theory we require two Grassmann coordinates.

The plan of this paper is as follows. In sec. II, we discuss the preliminaries about CS theory with its supersymmetric BRST invariance. Further, in Sec. II, we 
demonstrate the extended BRST invariant theory (including shift symmetry)  where
antifields gets their identifications naturally. The extended BRST invariant
superspace formulation of the theory is discussed in Sec. IV.
Sec. V is devoted to study the extended anti-BRST symmetry of the theory.
In sec. VI, we analyses both extended BRST as well as extended anti-BRST invariant
CS theory in superspace. We summaries our results in the last section.
\section{The CS theory and its BRST invariance}
In this section, we discuss the preliminaries of CS theory with
its BRST invariance. In this view, 
the CS term in $(2+1)$ flat  spacetime dimensions
is given by the following gauge invariant Lagrangian density
\begin{eqnarray}
{\cal L}_{CS} =-\mbox{Tr} \left[\frac{k}{4\pi} \epsilon^{\mu\nu\rho}\left(A_\mu\partial_\nu
A_\rho -\frac{2i}{3}A_\mu A_\nu A_\rho \right)\right],\label{cs}
\end{eqnarray}
where the inverse of the coupling constant $k$ is an integer
and $A_\mu$ is a Lie algebra valued gauge field, the corresponding group being chosen to be 
simple and compact.
The importance
  of the Chern-Simons action  (\ref{cs}) lies in the fact that  being the integral of
a 3-form it does not depend on the metric   which one can introduce on three dimensional manifolds. This important feature, therefore, allows to compute topological invariants of three
dimensional manifolds by using perturbative field theory techniques \cite{ew}.
The topological
character of the Lagrangian density (\ref{cs}) is the origin of the ultraviolet finiteness of 
the
perturbative Feynman diagrams expansion. This Lagrangian density yields a vertex functional   
which obeys the Callan-Symanzik equation with vanishing
$\beta$-function and no anomalous dimensions \cite{fd1,abl}.
This Lagrangian density possesses 
 following infinitesimal gauge invariance:
 \begin{eqnarray}
 \delta A_\mu =D_\mu\theta =\partial_\mu  \lambda +i[\lambda, A_\mu],
 \end{eqnarray}
$\lambda$ is  a Lie algebra valued local parameter. 
In order to fix the redundancy of gauge freedom in  the CS theory due to above gauge 
invariance 
(\ref{cs}), although there are many choices but we adopted the
axial gauge
\begin{equation}
n^\mu A_\mu =0,
\end{equation}
where $n^\mu$ is an arbitrary constant vector.

The gauge restriction can be achieved at quantum level by adding
following  gauge-fixing  and corresponding ghost terms in the
CS action (\ref{cs}):
\begin{eqnarray}
{\cal L}_{gf} &= & \mbox{Tr}  \left(Bn^\mu A_\mu  \right),\nonumber\\
{\cal L}_{gh} &= &-\mbox{Tr}  \left(\bar C n^\mu D_\mu C\right),
\end{eqnarray}
where $C$ and $\bar C$ are Faddev-Popov ghost and anti-ghost fields respectively.

Now, the total Lagrangian density for CS theory in axial gauge is given by
\begin{equation}
{\cal L}={\cal L}_{CS} +{\cal L}_{gf} +{\cal L}_{gh}, \label{tot}
\end{equation}
which is invariant under following nilpotent BRST transformations:
\begin{eqnarray}
s A_\mu &=&D_\mu C =\partial_\mu C +i[c,A_\mu],\nonumber\\
s C &=& iC^2,\nonumber\\
s\bar C&=&B,\nonumber\\
s B&=&0.\label{brs}
\end{eqnarray}
  Where $C^2$ is defined as
\begin{eqnarray}
 C^2 & \equiv & i f^{abc} C^b C^c  ,\nonumber
\end{eqnarray}
The combination of gauge-fixing and ghost terms is
BRST exact and, hence, can be written in terms of BRST variation of
gauge-fixing fermion  $\Psi = \left(\eta^{\mu\nu} \bar C n_\mu A_\nu\right)$ as follows
\begin{equation}
{\cal L}_{gf} +{\cal L}_{gh}=\mbox{Tr} [ s\left(\eta^{\mu\nu} \bar C n_\mu A_\nu\right)],
\label{ps}
\end{equation}
where $\eta^{\mu\nu}$ is the Minkowski metric of the 3-dimensional flat
spacetime.

However, for the Landau (covariant) gauge choice the 
gauge-fixing and ghost terms have the following expressions:
\begin{eqnarray}
{\cal L}'_{gf} &= & \mbox{Tr}  \left(B\partial^\mu A_\mu  \right),\nonumber\\
{\cal L}'_{gh} &= &-\mbox{Tr}  \left(\bar C \partial^\mu D_\mu C\right),
\end{eqnarray}
Employing these terms, the total Lagrangian density in Landau gauge reads
\begin{equation}
{\cal L}'={\cal L}_{CS} +{\cal L}'_{gf} +{\cal L}'_{gh}={\cal L}_{CS} +\mbox{Tr} [ s
\left(\eta^{\mu\nu} \bar C \partial_\mu A_\nu\right)], \label{tot1}
\end{equation}
which also remains invariant under the same set of BRST transformation (\ref{brs}).
\section{Extended BRST Invariant Lagrangian Density}
In this section, we analyses the extended BRST transformations for  CS theory
in BV formulation. The advantage of doing so is that antifields 
get identification naturally. We start the analysis by shifting  all the fields
from their original value as follows
\begin{equation}
A_\mu \longrightarrow  A_\mu- \tilde A_\mu \quad
C \longrightarrow  C - \tilde C \quad
\bar C \longrightarrow \bar C - \tilde {\bar C} \quad
B \longrightarrow B - \tilde B.
\end{equation}
Under such shifting of fields the Lagrangian densities (\ref{tot}) and (\ref{tot1}) also get 
shifted respectively as follows:
\begin{eqnarray}
\tilde{\cal L }&=& {\cal L}(A_\mu - \tilde A_\mu, C - \tilde C, \bar C - \bar C, B - \tilde 
B),\label{til}\\
\tilde{\cal L }'&=& {\cal L}'(A_\mu - \tilde A_\mu, C - \tilde C, \bar C - \bar C, B - \tilde 
B).\label{til1}
\end{eqnarray}
These shifted Lagrangian densities remain invariant 
under BRST transformation  in tandem with shift symmetry transformation, commonly
known as extended BRST transformation.
The Lagrangian densities (\ref{til}) and (\ref{til1}) admit the following extended BRST 
symmetry transformations:
\begin{eqnarray}
s A_\mu &=&\psi_\mu ,\nonumber\\
s \tilde A_\mu &=&\psi_\mu - D_\mu (C - \tilde C), \nonumber\\
s C &=& \epsilon  ,\nonumber\\
s \tilde C &=& \epsilon - i(C-\tilde C)^2, \nonumber\\
s \bar C &=& \bar \epsilon , \nonumber\\
s \tilde {\bar C} &=& \bar \epsilon - (B-\tilde B),\nonumber\\
s B &=& \rho ,\nonumber\\
s \tilde B  &=& \rho,\label{ex}
\end{eqnarray}
where $\psi_\mu, \epsilon, \bar \epsilon$ and $\rho$ are the ghost fields 
associated with shift
 symmetry for $A_\mu , C ,\tilde C$ and $B$ respectively.
To preserve the nilpotency of extended BRST symmetry (\ref{ex}) 
the ghost fields are required to have following BRST transformation
\begin{eqnarray}
s \psi &=& 0 , \nonumber\\
s \epsilon &=& 0 , \nonumber\\
s \bar \epsilon &=& 0 , \nonumber\\
s \rho  &=& 0.
\end{eqnarray}
We further need to introduce the anti-fields for the ghost fields $A^\star_\mu, C^\star, \bar 
C^\star$ and $B^\star$
in the theory
to make it ghost free. 
The BRST variation of anti-ghost fields are given by
\begin{eqnarray}
s A^\star_\mu &=& -\zeta_\mu , \nonumber\\
s C^\star &=& - \sigma, \nonumber\\
s \bar C^\star &=& -\bar \sigma , \nonumber\\
s B^\star &=& -\bar \upsilon,
\end{eqnarray} 
where $\zeta_\mu, \sigma , \bar \sigma $ and $\bar \upsilon$ are the  Nakanishi-Lautrup type 
auxiliary  fields 
corresponding to shifted fields $\tilde A_\mu, \tilde C, \tilde {\bar C}$ and $\tilde B$ with 
vanishing BRST variations
\begin{eqnarray}
s \zeta_\mu &=& 0 , \nonumber\\
s \sigma  &=& 0 , \nonumber\\
s \bar \sigma  &=& 0 , \nonumber\\
s \bar \upsilon &=& 0. 
\end{eqnarray}
Now, our original theory can be recovered by fixing the gauge  of shift 
symmetry properly such that all the tilde fields vanish.
We achieve this by adding following gauge-fixed Lagrangian density
in the shifted Lagrangian densities (\ref{til}) and (\ref{til1}):
\begin{eqnarray}
\tilde{\cal L}_{gf+gh} &=& \mbox{Tr}\left[-\zeta^\mu \tilde A_\mu - A_\mu^\star [\psi^\mu - D^
\mu (C - \tilde C)] - \sigma  \tilde {\bar C }
+ C^\star[\bar\epsilon  - (B - \tilde B)]\right.\nonumber\\
& -&\left.\bar \sigma  \tilde C + \bar C^\star[\epsilon  - i(C - \tilde C)^2] - \bar \upsilon 
\tilde B - B^\star \rho\right].\label{i}
\end{eqnarray}
The Lagrangian density $\tilde{\cal L}_{gf+gh}$ is also invariant under the extended BRST 
symmetry transformations mentioned above.
Now, performing equations of motion of auxiliary fields in the
above expression we obtain
\begin{eqnarray}
\tilde{\cal L}_{gf+gh} =\mbox{Tr}\left[  - A_\mu^\star (\psi^\mu - D^\mu  C)  
+ C^\star(\bar\epsilon  -  B  )+  \bar C^\star(\epsilon  - i C^2) - B^\star \rho\right].
\label{h}
\end{eqnarray}
The gauge-fixing and ghost terms of Lagrangian density are BRST exact
and can be expressed   in terms of a general gauge-fixing fermion 
$\Psi$ as 
\begin{eqnarray}
 \mbox{Tr}( s \Psi)&=&\mbox{Tr}\left[s A_\mu \frac{\delta \Psi}{\delta A_\mu} + s C 
 \frac{\delta \Psi}{\delta C} 
+ s \bar C \frac{\delta \Psi}{\delta \bar C} +  s B \frac{\delta \Psi}{\delta B}\right], 
\nonumber\\
&= &\mbox{Tr}\left[- \frac{\delta \Psi}{\delta A_\mu}\psi _\mu + \frac{\delta \Psi}{\delta C} 
\epsilon
+ \frac{\delta \Psi}{\delta \bar C}\bar\epsilon - \frac{\delta \Psi}{\delta B} \rho
\right],
\label{g}
\end{eqnarray}
  The Lagrangian densities in equations (\ref{til}), (\ref{h}) and (\ref{g}) together 
  describes the complete effective action for CS theory in axial gauge possessing extended
  BRST symmetry  as
\begin{eqnarray}
{\cal L}_{eff} & = & \tilde {\cal L} + {\cal L}_{gf}+{\cal L}_{gh} 
+\tilde{\cal L}_{gf+gh},\nonumber\\
  &=& \tilde {\cal L} +\mbox{Tr}\left[ \left(- A_\mu^\star - \frac{\delta \Psi}{\delta A^\mu}
  \right)\psi ^\mu +
  \left(\bar C^\star + \frac{\delta \Psi}{\delta C}\right) \epsilon +
  \left(C^\star + \frac{\delta \Psi}{\delta \bar C}\right)\bar\epsilon\right. \nonumber\\
 &+&\left.  \left(- B^\star - \frac{\delta \Psi}{\delta B}\right) \rho
  + A_\mu^\star D^\mu  C    -C^\star B   - i\bar C^\star  C^2\right],\label{eff}
\end{eqnarray}
where $\Psi$ refers here the gauge-fixing fermion corresponding to
the axial gauge.
Using equations of motion of the  ghost fields associated with shift symmetry, we obtain
\begin{eqnarray}
A_\mu ^\star &=& \frac{\delta \Psi}{\delta A^\mu} , \nonumber\\
\bar C^\star &=& - \frac{\delta \Psi}{\delta C}  , \nonumber\\
C^\star &=& - \frac{\delta \Psi}{\delta \bar C}  , \nonumber\\
B^\star &=& \frac{\delta \Psi}{\delta B}.
\end{eqnarray}
For the gauge-fixing fermion $\Psi$ corresponding to axial gauge  given in (\ref{ps}), the 
above expressions of anti-ghost fields yield
\begin{eqnarray}
A_\mu ^\star &=&    \eta_{\mu\nu} \bar C n^\nu, \nonumber\\
\bar C^\star &=& 0, \nonumber\\
C^\star &=& \eta^{\mu\nu}  n_\mu A_\nu, \nonumber\\
B^\star &=& 0.
\end{eqnarray}
Plugging these expression of anti-ghost fields in (\ref{eff}), we recover the Lagrangian 
density of our original CS theory in axial gauge.

However, for the gauge-fixing fermion corresponding to the Landau gauge given in (\ref{tot1}), 
the  anti-ghost fields 
\begin{eqnarray}
A_\mu ^\star &=&   - \eta_{\mu\nu}\partial^\nu \bar C, \nonumber\\
\bar C^\star &=& 0, \nonumber\\
C^\star &=& \eta^{\mu\nu}  \partial_\mu A_\nu, \nonumber\\
B^\star &=& 0.
\end{eqnarray}
With these values of anti-ghost fields, we can 
recover the Lagrangian density of  CS theory in Landau gauge.
\section{Extended BRST invariant superspace description}
In this section, we study the extended BRST invariant CS theory
in a superspace labelled by the coordinates $(x, \theta)$ where $\theta$ is Grassmann in 
nature and $x_\mu$ is space time in 2+1 dimension.
Superspace description for the extended BRST invariant theory is obtained
by defining the superfields of the form:
\begin{eqnarray}
A_\mu (x,\theta ) &=& A_\mu + \theta \psi_ \mu , \nonumber\\
\tilde A_\mu (x,\theta ) &=& \tilde A_\mu + \theta [\psi_ \mu - D_\mu(C - \tilde C)] , 
\nonumber\\
\chi (x,\theta ) &=& C + \theta \epsilon , \nonumber\\
\tilde \chi (x,\theta )&=& \tilde C + \theta [\epsilon - i(C - \tilde C)^2] , \nonumber\\
\bar \chi (x,\theta ) &=& \bar C + \theta \bar \epsilon , \nonumber\\
\tilde {\bar \chi} (x,\theta ) &=& \tilde {\bar C} + \theta [\bar \epsilon - (B - \tilde B) , 
\nonumber\\
B (x,\theta ) &=& B + \theta \rho , \nonumber\\
\tilde B (x,\theta ) &=& \tilde B + \theta \rho.
\end{eqnarray}
On the other hand, the super-antifields utilizing the extended BRST transformation for 
antifields are defined by
\begin{eqnarray}
\tilde A_\mu^\star(x,\theta ) &=& A_\mu^\star - \theta \zeta_\mu , \nonumber\\
\tilde \chi^\star (x,\theta ) &=& C^\star - \theta  \sigma       , \nonumber\\
\tilde {\bar \chi}^\star (x,\theta )&=& \bar C^\star - \theta \bar \sigma , \nonumber\\
\tilde B^\star (x,\theta ) &=& B^\star - \theta \bar \upsilon.
\end{eqnarray}
With the help of  these superfields and super-antifields, we  calculate 
\begin{eqnarray}
\frac{\delta(\tilde A_\mu^\star \tilde A^\mu)}{\delta \theta} &=& -A_\mu^\star[\psi^\mu - D^
\mu (C - \tilde C)] - \zeta_\mu \tilde A^\mu , \nonumber\\
\frac{\delta (\tilde {\bar\chi} ^\star \tilde \chi)}{\delta \theta} &=& \bar C^\star[\epsilon - i(C 
- \tilde C)^2] - \bar \sigma  \tilde C , \nonumber\\
\frac{\delta (\tilde {\bar \chi} \tilde \chi^\star)}{\delta \theta} &=& -\sigma \tilde {\bar 
C}  + C^\star[ \bar \epsilon - (B - \tilde B)], \nonumber\\
\frac{\delta (\tilde B^\star \tilde B)}{\delta \theta} &=&-B^\star \rho - \bar \upsilon \tilde 
B. \label{q}
\end{eqnarray}
Combining all the equations of (\ref{q}), we find that
\begin{eqnarray}
\mbox{Tr}\left[\frac{\delta}{\delta \theta} (\tilde A_\mu^\star \tilde A^\mu + \tilde {\bar 
\chi^\star} \tilde \chi + \tilde {\bar \chi} \tilde \chi^\star + \tilde B^\star \tilde B)
\right]&=&\mbox{Tr}\left[-A_\mu^\star[\psi^\mu - D^\mu (C - \tilde C)] - \zeta_\mu \tilde A^
\mu +\bar C^\star[\epsilon - i(C - \tilde C)^2]\right. \nonumber\\
&-& \left.\bar \sigma \tilde C 
-\sigma\tilde {\bar C}  + C^\star [\bar \epsilon - (B - \tilde B)]-B^\star \rho - \bar 
\upsilon \tilde B\right],
\end{eqnarray}
which is nothing but the shifted gauge-fixed Lagrangian density $\tilde{{
\cal L}}_{gf+gh}$ given in (\ref{i}).
Now, we define the general  super-gauge-fixing fermion written in superspace formulation  as 
follows
\begin{eqnarray}
\Phi(x, \theta) = \Psi (x)+ \theta (s\Psi),
\end{eqnarray}
which can further be expressed as
\begin{eqnarray}
\Phi(x, \theta) = \Psi (x)+ \theta \left(- \frac{\delta \Psi}{\delta A_\mu}\Psi _\mu + 
\frac{\delta \Psi}{\delta C} \epsilon
+ \frac{\delta \Psi}{\delta \bar C}\bar\epsilon - \frac{\delta \Psi}{\delta B} \rho\right).
\end{eqnarray}
So, the original gauge-fixing Lagrangian density in the superspace  can be defined as the left 
derivation of super-gauge-fixing fermion with respect to $\theta$ 
as $\mbox{Tr}\left[ \frac{\delta \Phi(x, \theta) }{\delta \theta}\right]$.

Hence, the complete effective action for the CS theory in general gauge  in the superspace is now 
given by
\begin{eqnarray}
{\cal L}_{gen} &=& \tilde {\cal L} +  \mbox{Tr}\left[\frac{\delta}{\delta \theta} (\tilde A_
\mu^\star \tilde A^\mu + \tilde {\bar \chi^\star} \tilde \chi + \tilde {\bar \chi} \tilde 
\chi^\star + \tilde B^\star \tilde B+\Phi)\right].
\end{eqnarray}
This compact expression indicates that the BV action of the extended CS theory in superspace 
is invariant under extended BRST transformations.
\section{Extended Anti-BRST Lagrangian Density}
In this section, we construct
the extended anti-BRST transformation under which the extended Lagrangian density remains 
invariant as follows 
\begin{eqnarray}
\bar s A_\mu &=&  A_\mu^\star + D_\mu (\bar C - \tilde {\bar C} ), \nonumber\\ 
\bar s \tilde A_\mu &=&  A_\mu^\star, \nonumber\\ 
\bar s C &=& C^\star + (B - \tilde B), \nonumber\\ 
\bar s \tilde C &=& C^\star, \nonumber\\ 
\bar s \bar C &=& (\bar C^\star - \tilde {\bar C}^\star) + i(\bar C - \tilde {\bar C})^2   , 
\nonumber\\ 
\bar s \tilde B &=& B^\star, \nonumber\\ 
\bar s B &=& B^\star.
\end{eqnarray}
The ghost fields associated with the shift symmetry  under extended anti-BRST symmetry
transforms as
\begin{eqnarray}
\bar s \psi_\mu  &=& \zeta_\mu , \nonumber\\ 
\bar s \epsilon  &=& \sigma , \nonumber\\ 
\bar s \bar \epsilon &=&  \bar \sigma  , \nonumber\\ 
\bar s \rho &=&   \bar \upsilon.                
\end{eqnarray}
However, under the extended anti-BRST transformation   the anti-fields of the auxiliary fields 
associated with the shift symmetry do not change
\begin{eqnarray}
\bar s \zeta_\mu = 0   ,\ \ \  \bar s A_\mu^\star = 0 ,\nonumber\\      
\bar s \sigma = 0  ,\ \  \ \bar s  C^\star = 0  ,  \nonumber\\ 
\bar s \bar\sigma= 0 ,\ \  \ \bar s \bar C^\star = 0,\nonumber\\ 
\bar s \bar \upsilon = 0,\ \ \ \bar s B^\star = 0.
\end{eqnarray}
The anti-gauge-fixing fermions 
for the CS 
theory in axial gauge ($\bar \Psi$) and in Landau gauge ($\bar \Psi'$) are respectively 
defined by
\begin{eqnarray}
\bar\Psi &= & \eta^{\mu\nu}  C n_\mu A_\nu,\nonumber\\
\bar\Psi' &= & \eta^{\mu\nu}  C \partial_\mu A_\nu.
\end{eqnarray}
The anti-BRST variation of  these gauge-fixing fermions  give the
corresponding gauge-fixing and ghost parts of the
complete Lagrangian density.
\section{Extended BRST and anti-BRST invariant superspace}
The extended BRST and anti-BRST invariant Lagrangian density is written in superspace with the
help of two Grassmannian coordinates $\theta$ and $\bar{\theta}$. Requiring the field strength 
to vanish along unphysical directions $\theta$ and $\bar{\theta}$
direction we determine the superfields in the following forms
\begin{eqnarray}
A_\mu (x,\theta , \bar \theta) &=& A_\mu (x) + \theta \psi_ \mu + \bar \theta[A_\mu^\star + D_
\mu \bar C] + \theta \bar \theta [\zeta_\mu + \partial_\mu \bar \epsilon]   , \nonumber\\
\tilde A_\mu (x,\theta , \bar \theta) &=& \tilde A_\mu (x) + \theta [\psi_ \mu - D_\mu(C-
\tilde C)] + \bar \theta A_\mu^\star + \theta \bar \theta \zeta_\mu , \nonumber\\
  \chi(x,\theta , \bar \theta) &=& C (x)+ \theta \epsilon + \bar \theta[C^\star + (B-\tilde 
  B)] +  \theta \bar \theta\sigma, \nonumber\\
 \tilde \chi(x,\theta , \bar \theta) &=& \tilde C(x) + \theta[\epsilon - iCC] + \bar \theta  
 C^\star + \theta \bar \theta \sigma, \nonumber\\
  \bar \chi(x,\theta , \bar \theta) &=&  \bar C(x) + \theta \bar \epsilon + \bar \theta[\bar 
  C^\star + i\bar C \bar C] + \theta \bar \theta \bar \sigma, \nonumber\\
 \tilde {\bar \chi}(x,\theta , \bar \theta) &=& \tilde {\bar C}(x) + \theta[\bar \epsilon - B] 
 + \bar \theta \bar C^\star + \theta \bar \theta \bar \sigma, \nonumber\\
B (x,\theta , \bar \theta) &=& B (x)+ \theta \rho + \bar \theta B^\star + \theta \bar \theta 
\bar \upsilon, \nonumber\\
\tilde B (x,\theta , \bar \theta) &=& \tilde B (x)+ \theta \rho + \bar \theta B^\star + \theta 
\bar \theta \bar\upsilon.
\end{eqnarray}
With these expressions of superfields we are able to establish the following relation
\begin{eqnarray}
-\frac { 1}{2} \mbox{Tr}\left[\frac {\partial}{\partial \bar \theta} \frac {\partial}{\partial 
\theta} (\tilde A_\mu \tilde A^\mu + \tilde \chi  \tilde {\bar \chi} + \tilde B  \tilde B)
\right] 
  &=& \mbox{Tr}\left[-\zeta^\mu \tilde A_\mu - A_\mu^\star[\psi^\mu - D^\mu(C - \tilde C)] - 
  \sigma \tilde {\bar C} + C^\star[\bar \epsilon - (B - \tilde B)]\right.\nonumber\\
  &- & \left.  \bar \sigma \tilde C +\bar C^\star[\epsilon - i(C - \tilde C)^2] - \bar 
  \upsilon \tilde B - B^\star \rho \right], \nonumber\\
&=&   \tilde {\cal L}_{gf+gh},
\end{eqnarray}
which is nothing but the shifted gauge-fixed Lagrangian density.
Being the $\theta \bar \theta$ component of a super field, this gauge-fixed Lagrangian density 
is manifestly invariant under the extended BRST and the anti-BRST transformations. 

Now, we define the general super-gauge-fixing fermion for the extended BRST and  the anti-BRST  
invariant theory
as follows
 \begin{eqnarray}
 \Phi(x,\theta , \bar \theta) &=& \Psi(x) + \theta (s \Psi) + \bar \theta (\bar s \Psi) + 
 \theta \bar \theta (s \bar s \Psi),
\end{eqnarray}
which yields the original gauge-fixing and ghost part of the complete
effective Lagrangian density as $\mbox{Tr}\left[\frac {\partial}{\partial \theta}\left[s(\bar 
\theta ) \Phi(x,\theta , \bar \theta)\right]\right]$.

Therefore, the complete Lagrangian density for the extended BRST and anti-BRST  invariant
CS theory in the general gauge can now be given by
\begin{eqnarray}
{\cal L}_{gen} =
 &=& \tilde {\cal L} -\frac { 1}{2} \mbox{Tr}\left[\frac {\partial}{\partial \bar \theta} 
 \frac {\partial}{\partial \theta} (\tilde A_\mu \tilde A^\mu + \tilde \chi  \tilde {\bar 
 \chi} + \tilde B  \tilde B)\right] + \mbox{Tr}\left[\frac {\partial}{\partial \theta}
 \left[s(\bar \theta ) \Phi(x,\theta , \bar \theta)\right]\right].
\end{eqnarray}
Performing equations of motion of auxiliary fields the shift fields can be removed from the 
above expression and by
integrating out the ghost fields for the shift symmetry we obtain the exact expressions
of antifields.

\section{Conclusion}
The $(2+1)$ dimensional CS theory  is subject of current 
interest because  of its some intriguing properties. For example,  the Green functions for the  
the model in axial gauge are shown the   unique and exact solution of the Ward identities 
without 
reference to any action principle \cite{sps}.  It is also well-known  that in axial gauge the 
the Faddeev-Popov determinant of this gauge-fixing
procedure is a constant function \cite{jf}.
 
In this work we have considered $(2+1)$ dimensional CS  theory in both the axial and the 
Landau gauges and 
have attempted to describe the extended BRST and anti-BRST 
invariant (including some shift symmetry) CS  theory in BV formulation. 
We show that antifields arises naturally in such 
formulation. We have further  
provided superspace and superfield description of such CS theory. We have shown that the BV 
action for such CS theory can be written in a manifestly extended BRST invariant manner in a 
superspace
with one fermionic coordinate. 
 However, a superspace with two Grassmann coordinates are required for a manifestly covariant 
 formulation 
of the extended BRST and extended anti-BRST invariant BV actions for CS gauge theory
in any arbitrary gauge.
It will be interesting to extend this formulation for anomalous gauge theories. 
\section*{Acknowledgments}
One of us (SU) acknowledges the support from the CNPq-Brazil under the Grant No. 
150155/2014-0.


\begin{thebibliography}{99}
\bibitem{as} A. S. Schwarz, Baku International Topological Conference, Abstracts, vol. 2, p. 
345, (1987).
\bibitem{ew}  E. Witten, Commun. Math. Phys. 121, 351 (1989).
\bibitem{ew1}  E. Witten, Nucl. Phys. B 322, 629 (1989).
\bibitem{town} A. Ach\'{u}carro and P. K. Townsend,  Phys. Lett. B 180, 9 (1986).
\bibitem{jf}  J. Fr\"{o}hlich and C. King, Commun. Math. Phys. 126167 (1989).
 
\bibitem{db}  D. Birmingham, M. Rakowski and G. Thompson, Nucl. Phys. B 329, 83 (1990). 
\bibitem{fd}  F. Delduc, F. Gieres and S. P. Sorella, Phys. Lett. B 225, 367 (1989).
\bibitem{fd1}  F. Delduc, C. Lucchesi, O. Piguet and S.P. Sorella, Nucl. Phys. B 346, 313 
(1990).
\bibitem{ab}  A. Blasi, O. Piguet, S. P. Sorella, Nucl. Phys. B 356, 154 (1991).

\bibitem{kla} K. Landsteiner, M. Langer, M. Schweda and S. P. Sorella, Phys. Lett. B 337, 294 
(1994).
\bibitem{eg}  E. Guadagnini, N. Maggiore and S.P. Sorella, Phys. Lett. B 255, 65 (1991).
\bibitem{bv} I. A. Batalin and G. A. Vilkovisky,   Phys.
Lett. B 102, 27 (1981).
\bibitem{bv1} I. A. Batalin and G. A. Vilkovisky, Phys. Rev. D 28, 2567 (1983) (Erratum-ibid.
D 30, 508 (1984))
\bibitem{wei} S. Weinberg,  The quantum theory of fields, Vol-II: Modern
applications, Cambridge, UK Univ. Press, 1996.
\bibitem{zinn} J. Zinn-Justin, Renormalization of Gauge Theories,   Springer-Verlag Berlin 
1975.
\bibitem{brs} C. Becchi, A. Rouet, R. Stora,  
Commun. Math. Phys. 42, 127 (1975).
\bibitem{brs1} C. Becchi, A. Rouet, R. Stora,  Annals Phys. 98,  287 (1976).
\bibitem {jm} S.D. Joglekar and B. P. Mandal, Phys. Rev. {D 49},  5617 (1994).  
\bibitem {jm1} S.D. Joglekar and B. P. Mandal, Phys. Rev. {D 55},  5038 (1997); Erratum ibid 
{\bf D59}
 (1999) 129902. 
\bibitem {jm2} S.D. Joglekar and B. P. Mandal, Z. Phys. {C 70},  673 (1996).
\bibitem{ad} J. Alfaro and P. H. Damgaard, {  Phys. Lett.} {B 222}, 425 (1989).
\bibitem{ad1} J. Alfaro, P. H. Damgaard , J. I. Latorre and D. Montano, {Phys. Lett.} {B 233}, 
153 (1989).
\bibitem{ad2} J. Alfaro and P. H. Damgaard, {Nucl. Phys.} {B 404}, 751 (1993).
\bibitem{ba} N. R.F. Braga and A. Das, {Nucl. Phys.} {B 442}, 655 (1995).
\bibitem{fk} M. Faizal and M. Khan, {Eur. Phys. J.} {C 71}, 1603 (2011).
\bibitem{subm} S. Upadhyay and B. P. Mandal, Eur. Phys. J. C 72, 2059 (2012).
\bibitem{sudh} S. Upadhyay, in preparation.
\bibitem{brax} P. Brax and J. Martin, Phys. Rev. D 72, 023518 (2005).
\bibitem{heb} A. Hebecker, A. K. Knochel and T. Weigand,   JHEP 1206, 093 (2012).
\bibitem{sud} S. Upadhyay, Phys. Lett. B 723,  470 (2013).
 
\bibitem{abl} A. Blasi and R. Collina, Nucl. Phys. B 345, 472 (1990).
\bibitem{sps} A. Brandhuber, S. Emery, M. Langer, O. Piguet, M. Schweda and S.P. Sorella, 
Helv. Phys. Acta 66,  551 (1993).

\end{thebibliography}
\end{document}